\documentclass[onecolumn,prl]{revtex4}
\usepackage{bm}
\usepackage{epsfig}
\usepackage{amsmath}
\begin{document}
\title{Towards Boltzmann Distribution}
\author{Aniruddha Chakraborty \\
School of Basic Sciences, Indian Institute of Technology Mandi,\\
Mandi, Himachal Pradesh, 175001, India}
\date{\today }
\begin{abstract}
The Boltzmann distribution (the most probable distribution) is one of the most important concepts used in physics, chemistry and biology. Suppose we put the system initially in one of the less probable state then the system will find the most probable state by a random search among all possible energy distributions and thus can take long time depending on the size of the system. In the following, simple analysis using our simple model shows that a small and physically reasonable energy bias against locally unfavorable energy distribution, of the order of a few $kT$, can reduce the time-scale of the process by a significant size.
\end{abstract}
\maketitle
The Boltzmann distribution is one of the most important concepts used in physics, chemistry and biology \cite{Davidson,Chandler,Atkins,McQuarrie}. A basic statement of the Boltzmann distribution in our context is, in a system of particles, the probability of a particle being in a state with energy
$\epsilon$ is given by
\begin{equation}
P(\epsilon_i)= g(\epsilon_i)e^{- \beta \epsilon_i}
\end{equation}
where $g(\epsilon_i)$ is the degeneracy of energy $\epsilon_i$ and $\beta$ is a constant which determines the average energy.
Also, $\beta^{-1}=kT$, where $k$ is Boltzmann's constant. In this paper, we assume the system is composed of a fixed number of particles denoted by
$n_{tot}$. Each particle is described by one or more one-particle quantum numbers, $i$, and a one-particle quantum energy, $\epsilon_i$, a function of the quantum numbers. Particles need not be identical, but the total energy of the system is given, to a good approximation, by
\begin{equation}
\epsilon_{total}=\sum_{i} n_{i} \epsilon_i
\end{equation}
where $n_i$ is the number of particles with energy $\epsilon_i$. We also assume the particles to be in dynamic equilibrium with sufficiently rapid exchange of energy through interactions between particles. For generality, we note that the quantum numbers, $i$, need not be one-particle quantum numbers; they could be collective excitations, e.g., phonons or normal modes. Suppose we put the system initially in one of the less probable state then the system will find the most probable state by a random search among all possible energy distributions and thus can take an enormously long time. In the following simple analysis using our simple model shows that a small and physically reasonable energy bias against locally unfavorable energy distribution, of the order of a few $kT$, can reduce the time-scale of the process by a significant size. We start with $n_{tot}$ number of particles and infinite number of energy levels, but only finite number of energy levels (say $N$) will be occupied. If $n_i$ is the number of particles having energy $\epsilon_i$, then this number $n_i$ can be characterized as "correct" or "incorrect", based on the final or perfect value $n_p$ (correct if $n_i = n_p$, incorrect if $n_i \not= n_p$). Correct number of particles are labeled as $c$, and incorrect number of particles are labeled $i$. A typical distribution of energy in terms number of particles in each of the total $N$ energy levels can be expressed by a list of length $N$ as ciciicicicciicc..... It is obvious that the final or completely correct distribution is the one consisting of all $c$'s and no $i$'s. Now the problem is, starting from any arbitrary distribution of correct and incorrect number of particles at different energy levels, find how long it takes to get to the perfect distribution of energy for the first time ?  We assume that a correct value of number of particles at different energy level can become incorrect ($c \rightarrow i$) with the rate $k_{ci}$ and an incorrect value of number of particles at different energy levels can become correct ($i \rightarrow c$) with the rate $k_{ic}$ and that these changes occur entirely independently. As a result, the number ($s$) of incorrect number of particles at different energy levels changes in time. The first-passage time to the final state is the time required, starting from some arbitrary initial state $s$, to reach at $s = 0$ for the first time. The mean first-passage time $\tau(s)$ is the average of this required time over all ways of getting from $s$ to $s = 0$. We start with $s$ incorrect number of particles in different energy levels, so the correct number of particles at different energy levels is $n-s$. The rate at which $s \rightarrow s+1$, is the correct number number of particles times the rate $k_{ci}$ of changing a correct number of particles to an incorrect number of particles.
\begin{equation}
Rate\;_{(s \rightarrow s+1)}= k_{ci} (n-s).
\end{equation}
Similarly, the rate at which $s \rightarrow s-1$ is the incorrect number of particles times the rate $k_{ic}$ of changing an incorrect number of particles to a correct number of particles.
\begin{equation}
Rate\;_{(s \rightarrow s-1)}= k_{ic} s.
\end{equation}
The probability that there are $s$ incorrect number of particles at time $t$ is denoted by $P(s,t)$. This changes by gains from $s - 1$ and $s
+ 1$ and losses to $s - 1$ and $s + 1$. The master equation is
\begin{equation}
\frac{d P(s,t)}{dt}= (n - s + 1) k_{ci} P( s - 1, t) + (s + 1) k_{ic} (s + 1, t) - (n - s) k_{ci} P(s, t) - s k_{ic} P(s, t),
\end{equation}
with the assumption that $P(-1, t)$ and $P(N + 1, t)$ are both equal to $0$. Now we will use the standard procedure for using a master equation to find
mean first-passage times. We start with the differential equations for $P$ in matrix form as
\begin{equation}
\frac{d P(s,t)}{dt}= \sum_{s'} w(s,s') P(s',t).
\end{equation}
Impose an absorbing boundary condition at $S = 0$, so that only the states $S = 1$ to $N$, are involved. Then the fundamental
equation that determines the mean first passage times is
\begin{equation}
\sum_{s_0} \tau(s_0) w(s_0,s) = - 1,\;\\;\ for\;\; all \;\; s.
\end{equation}
So that
\begin{equation}
s k_{ic} \left[\tau(s-1) - \tau(s)\right]+ (n-s) k_{ci} \left[\tau(s-1) - \tau(s)\right]= -1.
\end{equation}
for all $s$ between $1$ and $N$. It is understandable that $\tau(0)$ must vanish and $\tau(N + 1)$ is never required. This determines all the other
$\tau(s)$.
\begin{equation}
\tau(s)= \frac{1}{k_{ci}}(1+ k_{eq})^N k_{eq} \int_{0}^{1}dy \frac{1 - y^s}{1-y} (1+{k_{eq}} y)^{-N-1},
\end{equation}
where ${k_{eq}}=k_{ci}/k_{ic}$. For large $N$, the integral is dominated severely by the contribution from small values of $y$. It is very weakly dependent on $s$. Its asymptotic form for large $N$ is given by
\begin{equation}
\tau(s)= \frac{1}{k_{ci}}(1+ k_{eq})^N \left[1+ 1! (N {k_{eq}})^{-1}+ 2! (N {k_{eq}})^{-2}+.....\right]
\end{equation}
The $s$-dependent parts of $\tau$ are in general negligibly small in comparison with the other terms, so the approximate result is given by
\begin{equation}
\tau(s) \approx \frac{1}{n k_{ci}}\left(1 + k_{ci}/k_{ic} \right) n.
\end{equation}
This asymptotic approximation is not valid for very small values of $k_{ci}$. The time $\tau$ is now independent of the starting value of $s$, that is interesting because even if the starting configuration is close to final one, there is a significant probability that it will wander further away before reaching $s = 0$. The mean first-passage time for a fully biased search, where the change $c \rightarrow i$ is not allowed so that $k_{ci} = 0$, is given by
\begin{equation}
\tau(s)= \frac{1}{k_{ic}}\sum_{j=1}^{s} \frac{1}{j}.
\end{equation}
In this limit, $\tau$ is independent of $N$ and has a logarithmic dependence on $s$. Up to this point, the energy distribution was characterized only by $N$
and the two rate constants ($k_{ci}$ and $k_{ic}$). Now, it will be interesting if we understand the deeper meaning of the ratio ${k_{ci}}/{k_{ic}}$. The kinetics for number of particles in any energy level is given by
\begin{equation}
\frac{d}{dt} [c]= - k_{ci} [c]+ k_{ic}[i]\;\;\; with\;\; [c]+[i]=1.
\end{equation}
So the equilibrium constant is defined as
\begin{equation}
k_{eq}=\frac{k_{ci}}{k_{ic}}=\frac{[i]_{eq}}{[c]_{eq}}.
\end{equation}
So that 
\begin{eqnarray}
[c]_{eq}=\frac{1}{1+k_{eq}}\;\; and \\ \nonumber  
[i]_{eq}=\frac{k_{eq}}{1+k_{eq}}.
\end{eqnarray}
It is easy to understand the concept of equilibrium constant $k_{eq}$ from statistical physics. Suppose the free energy for correct number of particles in the $i$-th energy level is $G_c$, and free energy for incorrect number of particles in the same energy level  is given by $G_i = G_c + U$. So $U$ is an free energy increament for having an incorrect number of particles, $U$ is definitely a positive number but it magnitude depends on the extent of 'incorrectness'. Then by using the equilibrium statistical thermodynamics, one gets
\begin{equation}
k_{eq} = \frac{k_{ci}}{k_{ic}} = e^{- U / kT}
\end{equation}
So that $k_{ci}/k_{ic}$ is small, if $U$ is larger than $kT$ and $\tau$ can become much smaller. Also at higher temperature the system will attain Boltzmann distribution faster.

\end{document}